\documentclass[sigconf,screen]{acmart}
\copyrightyear{2026}
\acmYear{2026}
\setcopyright{cc}
\setcctype{by-nc-nd}
\acmConference[ASE '26]{Proceedings of the 41st IEEE/ACM International Conference on Automated Software Engineering}{October 12--16, 2026}{Munich, Germany}
\acmBooktitle{Proceedings of the 41st IEEE/ACM International Conference on Automated Software Engineering (ASE '26), October 12--16, 2026, Munich, Germany}
\acmDOI{10.1145/3832783.3837552}
\acmISBN{979-8-4007-2882-2/2026/10}
\acmSubmissionID{ase26main-p3524-p}
\received{2026-03-26}
\received[accepted]{2026-06-18}

\usepackage{microtype}
\usepackage{enumitem}
\usepackage{pifont} 
\usepackage[many]{tcolorbox}
\usepackage{cleveref}
\usepackage{xspace}
\usepackage{booktabs} 
\usepackage[table,xcdraw]{xcolor} 

\newcommand{\modify}[1]{\textcolor{black}{#1}}

\definecolor{revisionblue}{RGB}{0,0,200}

\DeclareRobustCommand{\mybox}[2][gray!20]{%
\begin{tcolorbox}[
        breakable,
        left=0pt,
        right=0pt,
        top=0pt,
        bottom=0pt,
        colback=#1,
        colframe=#1,
        width=\linewidth,
        enlarge left by=0mm,
        boxsep=5pt,
        arc=0pt,outer arc=0pt,
        ]
        #2
\end{tcolorbox}
}

\begin{document}

\title{Implicit, Yet Impactful: Understanding Hidden Dependencies in Java Projects}

\author{Lyuye Zhang}

\orcid{0000-0003-3087-9645}
\affiliation{%
  \department{College of Cryptology and Cyber Science}
  \institution{Nankai University}
  \city{Tianjin}
  \country{China}
}
\affiliation{%
  \institution{Nanyang Technological University}
  \city{Singapore}
  \country{Singapore}
}
\email{zh0004ye@e.ntu.edu.sg}

\author{Chengwei Liu}
\correspondingauthor
\authornote{Corresponding author.}
\orcid{0000-0003-1175-2753}
\affiliation{%
  \department{College of Cryptology and Cyber Science}
  \institution{Nankai University}
  \city{Tianjin}
  \country{China}
}
\email{chengwei.liu@nankai.edu.cn}

\author{Fangyuan Zhang}
\orcid{0009-0000-9599-1369}
\affiliation{%
  \institution{Nankai University}
  \city{Tianjin}
  \country{China}
}
\email{fangyuanzhang@mail.nankai.edu.cn}

\author{Yiran Zhang}
\orcid{0000-0002-9366-6076}
\affiliation{%
  \institution{Nanyang Technological University}
  \city{Singapore}
  \country{Singapore}
}
\email{yiran002@e.ntu.edu.sg}

\author{Yuan Zhou}
\orcid{0000-0002-1583-7570}
\affiliation{%
  \institution{Zhejiang Sci-Tech University}
  \city{Hangzhou}
  \country{China}
}
\email{yuanzhou@zstu.edu.cn}

\author{Yang Liu}
\orcid{0000-0001-7300-9215}
\affiliation{%
  \institution{Nanyang Technological University}
  \city{Singapore}
  \country{Singapore}
}
\email{yangliu@ntu.edu.sg}

\begin{abstract}
As software usage continues to expand, package managers automatically resolve dependencies to construct a dependency graph based on user-specified requirements. These explicitly declared dependencies, known as direct dependencies, receive significant attention in terms of maintainability and security. However, implicit dependencies, which are not explicitly defined by users but are still directly utilized or referenced in their project code due to oversight, remain largely unnoticed. Unlike ordinary transitive dependencies, which may remain unused and invisible to the root, implicit dependencies are actively used yet undeclared, leaving their versions outside the project’s direct control. This lack of awareness poses substantial challenges related to security and maintainability. 

In this study, we present the first study to treat implicit dependencies as the focal phenomenon and quantitatively characterize their lifecycle consequences for the Maven ecosystem.
\modify{We meticulously collected and built a large-scale dataset with 1,157 libraries with 19,812 versions from the Maven Central Repository and 972 modules from GitHub.}
Our findings reveal that 34.12\% of the analyzed dataset contains implicit dependencies, with two primary causes identified as key contributors to the issue. Among these, 48\% introduce breaking changes due to version drift, and 36 CVEs have vulnerable methods directly used by root projects; 30.28\% of implicit dependencies are affected by known vulnerabilities under the version-range convention SCA tools use for declared dependencies. Finally, we identified and analyzed four major countermeasures, providing actionable insights and practical implications for addressing this overlooked issue for stakeholders within the OSS ecosystem.
\end{abstract}



\maketitle

\section{Introduction}
\label{sec:intro}
As open-source software (OSS) reuse grows, effective dependency management has become crucial for reusing code while minimizing risk. Mature package managers such as Maven~\cite{maven}, NPM~\cite{npm}, and PyPI~\cite{pypi} let developers integrate dependencies without directly managing the transitive dependencies these rely on. Focusing on direct dependencies, developers pin stable versions and run Software Composition Analysis (SCA)~\cite{sca} or audit tools such as Dependabot~\cite{dependabot} for security and license checks; direct dependencies thus rightfully receive significant attention to security, maintainability, and stability.

Unfortunately, when transitive dependencies are directly referenced in a project’s code but not declared, they become implicit dependencies~\cite{cataldo2009software} (sometimes called ghost dependencies). 
Unlike ordinary transitives, which may remain unused and only serve the needs of other libraries, implicit dependencies are actively used by the root project yet omitted from its configuration, leaving their versions outside the developer’s explicit control. 
This situation often arises because package managers automatically download transitive libraries together with direct ones, making them appear indistinguishable during development. 
Although convenient, such reliance conflicts with best-practice guidelines (e.g., Google JLBP~\cite{jlbp}), which require that all code-referenced libraries be declared as direct dependencies, including implicit dependencies. 
Thus, implicit dependencies represent a critical subset of transitives that should have been promoted to direct dependencies but remain undeclared, creating risks of version drift, breaking changes, and overlooked vulnerabilities.

Implicit dependencies, lacking user-specified versions, are automatically resolved by package managers based on the project context, making their versions volatile. This phenomenon, known as version drift, can lead to breaking changes and compromise project stability. 
Additionally, many widely used SCA tools focus exclusively on direct dependencies, overlooking vulnerabilities in implicit dependencies. Examples include Dependabot~\cite{dependabot} (GitHub), Dependency Check~\cite{owaspcomponent} by OWASP~\cite{owasp}, DepShield~\cite{depshield} by Sonatype~\cite{sonatype}, NPM Audit~\cite{npmaudit} by NPM~\cite{npm}, Safety~\cite{safety} for Python, and Xray~\cite{jfrogxray} by JFrog~\cite{jfrog}. As a result, implicit dependencies can harbor hidden security vulnerabilities, posing risks equivalent to those of direct dependencies.

The concept of implicit dependency was introduced in the academic literature years ago~\cite{cataldo2009software} without in-depth examination.
Follow-up work has touched on the used-but-undeclared phenomenon without isolating it as the focus of a longitudinal, risk-oriented study: Soto-Valero et al.~\cite{soto2021comprehensive} identify an analogous \emph{used-transitive} category, but only as one of six labels in a study of dependency bloat. In NPM the phenomenon surfaces only as a secondary concern: Javan Jafari et al.~\cite{javanjafari2022dependency} flag a \emph{missing dependency} smell among several, and Latendresse et al.~\cite{latendresse2022not} note transitively-resolved missing peer dependencies but exclude them.
Given their significant and often overlooked consequences, implicit dependencies must be properly identified and managed to ensure the long-term sustainability of the OSS ecosystem.

To bridge this research gap, we present, to the best of our knowledge, the first study to treat implicit dependencies as the focal phenomenon and quantitatively characterize their diverse lifecycle consequences in the Maven ecosystem, spanning a longitudinal trend across 19,812 versions (Section~\ref{sec:timeline}), version-drift-induced breaking changes at method-reachability granularity (Section~\ref{sec:breakingcode}), method-reachable CVE exposure specific to the implicit subset (Section~\ref{sec:vuln}), and resolution outcomes traced through evolution (Section~\ref{sec:countermeasure}).
Specifically, we conducted a large-scale empirical study on Maven Central artifacts and actively developed GitHub projects. Using points-to analysis, we identified implicit dependencies and analyzed their timeline evolution, the impact of version drift (including breaking changes from passive updates), actually-invoked vulnerabilities, and how they are resolved in practice together with potential countermeasures. Finally, we provide actionable implications to raise awareness of implicit dependencies and promote their effective management.

We summarize our key contributions as follows:
\begin{itemize}[leftmargin=1.5em]
    \item \textbf{First study to treat implicit dependencies as the focal phenomenon.}
    While related work treats used-but-undeclared dependencies as one label within a smell taxonomy~\cite{soto2021comprehensive,javanjafari2022dependency,latendresse2022not}, we present the first large-scale analysis that isolates implicit dependencies and quantifies their lifecycle consequences, spanning $1,157$ Maven Central libraries and $972$ GitHub projects.

    \item \textbf{Longitudinal analysis of persistence and growth.} 
    By examining multiple project releases, we show that implicit dependencies not only persist but frequently increase over time. 
    This demonstrates that implicit dependencies are a long-term phenomenon rather than short-lived anomalies.

    \item \textbf{Quantification of risks through version drift and vulnerabilities.} 
    We measure how version drift in implicit dependencies leads to breaking changes, and we map implicit dependencies to known CVEs. Because they are undeclared, these maintenance and security risks are more likely to be overlooked by developers running dependency audits than those of directly declared dependencies, even though their magnitude is comparable.

    \item \textbf{Causes, countermeasures, and actionable guidance.}
    We analyze the underlying causes of implicit dependencies and categorize observed countermeasures.
    Based on these findings, we provide actionable recommendations for developers and usage-aware improvements for SCA tools.
\end{itemize}

\subsection{Motivating Examples}
\label{sec:motivating}

To illustrate the real-world impact of implicit dependencies, we present three representative examples drawn from our dataset, each highlighting a distinct risk category.

\noindent\textbf{Example 1: Hidden Vulnerability via Jackson.}
The JWT library \texttt{jjwt-jackson} implicitly depends on \texttt{jackson-core}---never declared in its \texttt{pom.xml} but directly used in code.
This implicit dependency spans 30 versions (2.9.6--2.12.7), every one affected by at least one CVE (e.g., CVE-2019-16943, CVSS 6.8, a deserialization flaw in \texttt{jackson-databind}).
Because \texttt{jackson-core} is undeclared, SCA tools may fail to flag the vulnerability.

\noindent\textbf{Example 2: Version Drift in Spring.}
\texttt{spring-retry} implicitly depends on \texttt{spring-core}, referencing its logging APIs (\texttt{LoggerFactory}, \texttt{Logger}).
The initial resolved version (3.0.5.RELEASE) carried 14 CVEs.
Over the project's lifetime, this implicit dependency drifted across 138 versions (up to 6.2.0) before maintainers eventually promoted it to a direct dependency---illustrating substantial technical debt that accumulated during the implicit phase.

\noindent\textbf{Example 3: Fragile Build in Data Platform.}
The Dinky data platform (\texttt{dinky-common:1.2.0}) uses \texttt{commons-codec} for Base64 and Hex encoding in authentication modules, yet never declares it.
The library arrives transitively at depth~2 among 142 transitive dependencies, so upgrading or removing an intermediate dependency could silently break encoding functionality across multiple modules (\texttt{dinky-alert-feishu}, \texttt{dinky-alert-dingtalk}).

These examples show that implicit dependencies affect diverse project types and manifest as security vulnerabilities, breaking changes, and fragile builds. The remainder of this paper systematically quantifies these phenomena.

\section{Background}
\label{sec:background}
\subsection{Maven Dependency Management}
\label{sec:versionresolution}

Maven is one of the most widely used build automation and dependency management tools. It simplifies dependency resolution by allowing developers to declare required libraries in \texttt{pom.xml} files. Maven Central Repository (MCR)~\cite{mvnrepo} organizes libraries using a GAV (Group ID, Artifact ID, Version) coordinate system~\cite{mvnartifact}. 
Maven simplifies the management by automatically resolving transitive dependencies. However, this process introduces version uncertainty caused by the diamond dependency (multiple required versions of one library). Maven’s \textit{nearest-first} strategy selects the version closest to the root without guaranteeing consistency across environments, leading to potential conflicts~\cite{wang2021will}.
These conflicts can result in runtime failures due to API incompatibilities and security risks from inadvertently including outdated or vulnerable versions.

\subsection{Term Definition}
\label{sec:termdef}
\begin{itemize}[leftmargin=1.5em]
    \item \textbf{Root project}: The general name of the target project to be analyzed from MCR and GitHub. 
    \item $MCR_{root}$: The sub-group of root projects from MCR.
    \item $GH_{root}$: The sub-group of root projects from GitHub.
    \item \textbf{Direct Dependencies}: Dependencies that are explicitly declared in the project’s build configuration (e.g., \texttt{pom.xml} in Maven). 
They represent libraries that the root project explicitly acknowledges and controls by specifying versions. 
Best-practice guidelines (e.g., Google JLBP~\cite{jlbp}) recommend declaring every dependency that is directly used in the project’s code as a direct dependency.
\item \textbf{Transitive Dependencies}: Dependencies that are pulled in indirectly through explicitly declared direct dependencies.
They may or may not be referenced in the root project's code. For example, if project $A$ depends on $B$, and $B$ depends on $C$, then $C$ is a transitive dependency of $A$.
Note that a dependency declared only in a parent POM is treated as a direct dependency.
\item \textbf{Implicit Dependencies}: A direct usage relationship in the root project's code (e.g., a call to a class, method, or field) that is satisfied by a transitively resolved artifact rather than by a dependency declared in the project's configuration (e.g., \texttt{pom.xml}). Implicit dependencies are thus a subset of direct dependencies.
Unlike ordinary transitives, which can remain unused and invisible to the root, implicit dependencies are actively used yet undeclared.
This notion coincides with what prior work has called \emph{used-but-undeclared} or \emph{used-transitive} dependencies~\cite{soto2021comprehensive,javanjafari2022dependency} (Section~\ref{sec:relatedwork}); we adopt the standard term ``implicit dependency''.
    
\end{itemize}

\section{Empirical Study}

\begin{figure*}[t]
  \centering
  \includegraphics[width=0.99\linewidth]{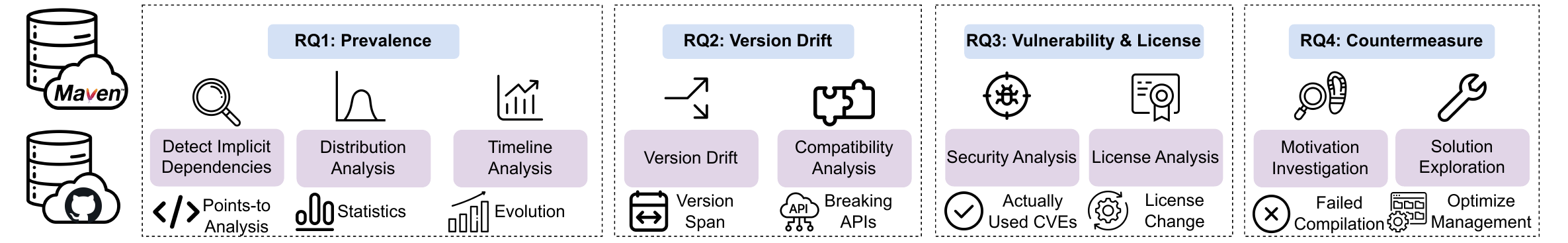}
  \caption{Overview of This Empirical Study}
  \label{fig:overview}
\end{figure*}

We aim to answer the following RQs as in Figure~\ref{fig:overview}:

\begin{itemize}[leftmargin=1.5em]
    \item \textbf{RQ1:} How prevalent are implicit dependencies in Maven projects, and why are they introduced?
    \item \textbf{RQ2:} How much version drift do implicit dependencies introduce, and what problems can version drift bring?
    \item \textbf{RQ3:} How many potential security vulnerabilities stem from implicit dependencies?
    \item \textbf{RQ4:} Are developers aware of these implicit dependencies, and how do they react to them?

\end{itemize}

\subsection{Initial Dataset Collection}

To fully understand the Java ecosystem and achieve generalizable conclusions, we analyze both the \textbf{published artifacts} ($MCR_{root}$) from MCR and the \textbf{in-development projects} ($GH_{root}$) hosted on GitHub repositories. The sampled, analyzed, and excluded artifacts are summarized in Table~\ref{tab:overview}. The two datasets are analyzed at a comparable artifact granularity but reach it differently (library for MCR and module for GitHub as one repository could host multiple modules), and their exclusions occur at respective stages. The repository level was not analyzed because the module is the unit actually built and, when published, scanned by SCA tools; aggregating to the repository level would hide cases where one module is clean while a sibling is not.

\subsubsection{\textbf{Maven Central Repositories}}
\label{sec:mcr}

To study implicit dependencies in widely used Java libraries, we first ranked artifacts in MCR based on their displayed usage statistics and selected the most used 1.2k libraries denoted as $MCR_{root}$, covering a total of 19,812 versions. For each version, we crawled \texttt{pom.xml} files and JAR artifacts associated with $MCR_{root}$. 43 of 1.2k were excluded due to the failure of crawling JAR files, resulting in 1,157 libraries for further analysis.
Dependency resolution in Maven is time-dependent: soft version constraints and repository updates may lead to different transitive versions being returned at different points in time.
To ensure replicability, all dependency trees in our study were resolved in May 2025 using Maven 3.9.4 on Java 17 with the official Maven Central mirror.
For each analyzed project version, we archived the complete dependency tree resolved by the effective POM and the corresponding JAR files in our dataset to ensure all version states are captured and locked against the May-2025 state of Maven Central. Section~\ref{sec:vulndata} states the same convention explicitly for the CVE-mapping step, where it matters most.
In spot checks where dependency resolution was repeated at different times, we did not observe variations for the same project version, but we acknowledge that, in principle, such variations may occur.
Therefore, we treat our collected dependency trees as a consistent snapshot of Maven Central as of May 2025 and released the raw data.

Since each library in the MCR may have multiple versions, sometimes numbering in the thousands, it is impractical to analyze all versions. Instead, our objective is to analyze the representative samples of MCR libraries.

To achieve this, we employed a version sampling strategy to ensure representative analysis while maintaining computational feasibility:
\ding{172} If a library contains 20 or fewer versions, all versions were included in the analysis.
\ding{173} If a library has more than 20 versions, we first sorted them based on semantic versioning~\cite{semver} order.
\ding{174} From the sorted versions, we then evenly sampled 20 versions, ensuring that the earliest and latest versions were always included.
\ding{175} Pre-release versions (e.g., \textit{snapshot, beta, alpha, RC}) were excluded from our sampling, as they often contain unstable features.
This sampling approach ensures that our study captures longitudinal trends in the implicit dependencies without being skewed by excessive version duplication within a single library.

For the transitive dependencies to be analyzed, since transitive dependencies are not explicitly listed in \texttt{pom.xml}, we extracted them using the \texttt{mvn dependency:\allowbreak tree} command. Additionally, we collected the JARs of these dependencies to analyze their actual usage in root libraries $MCR_{root}$. 

\subsubsection{\textbf{GitHub Java Repositories}}
\label{sec:ghrepos}
As a comparison, we also conducted the implicit dependency analysis in real-world popular GitHub projects.
Ranking stars in descending order, we collected 1,000 Java repositories from GitHub following the star-oriented data sampling strategy~\cite{soto2021comprehensive,wu2023understanding,zhang2023mitigating,zhang2023compatible}. One repository comprises multiple modules, which are analyzed individually as they are the same granularity as libraries. Among these, 368 projects were manually identified as valid Java projects and managed by Maven. 
The rest was manually excluded by the authors based on repository content: a repository was kept if it contained a top-level \texttt{pom.xml} describing a buildable Java artifact, and was excluded if it was a coding-interview-question collection, a tutorial or ``awesome-list'' aggregation, a course assignment repository, or otherwise did not represent a real, deployable software project; the per-repository inclusion/exclusion labels are released with our artifact~\cite{dataset}.
To avoid duplicated analysis of the projects hosted at both MCR and GitHub, 226 projects with GAV coordinates at MCR were ruled out, leaving 142 repositories.

For the 142 non-Maven repositories, as the compiled class files are mandatory for code-centric analysis, the repositories without compilable source code were further ruled out. A total of 48 repositories failed to compile completely by \texttt{mvn} either due to private dependencies or incompatible environments, leaving 94 remaining. Java projects could consist of multiple artifacts under one vendor, each of which is considered as an individual module with an optional standalone class file. Eventually, we obtained 972 modules with dependency trees and class files, serving as $GH_{root}$.

\modify{In total, we gathered 15,734 JAR files from MCR and GitHub after de-duplication.}

\begin{table}[t]
\footnotesize
\centering
\setlength{\tabcolsep}{2pt}
\caption{Dataset construction with analysis units, and implicit-dependency prevalence. Modules are the same granularity as libraries.}
\label{tab:overview}
\begin{tabular}{@{}lrrrr@{}}
\toprule
\rowcolor[HTML]{EFEFEF}
\textbf{Dataset (unit)} & \textbf{Sampled} & \textbf{Excluded} & \textbf{Analyzed} & \textbf{Implicit dependency\ (\%)} \\ \midrule
$MCR_{root}$ (lib) & 1{,}200 lib & 43 lib & 1{,}157 libs & 508 libs (43.90\%) \\
\rowcolor[HTML]{EFEFEF} $GH_{root}$ (module) & 1000 repo & 810+48 repo & 972 modules & 109 modules (11.21\%) \\
\bottomrule
\end{tabular}
\end{table}

\subsection{RQ1: Prevalence of Implicit Dependencies}
We aim to address \textbf{RQ1} by investigating the prevalence of implicit dependencies within \textbf{MCR} and \textbf{GitHub}. 

\subsubsection{\textbf{Methodology of Implicit Dependency Identification}}
\label{sec:method}
To quantify implicit dependencies, we employ \textbf{Points-to analysis}~\cite{pointeranalysis} to examine the transitive dependencies referenced in root projects at the bytecode level. Points-to analysis has been widely studied in Java static analysis \cite{tan2025interactive, roth2024axa, halalingaiah2024art, wimmer2024scaling, kozak2025skipflow}. We employ \textbf{SootUp}~\cite{sootup}, a well-established framework for bytecode analysis, due to its ability to extract program structures, track inter-class references, and analyze large-scale Java applications.

Our methodology follows the Java convention that a class is considered \textbf{referenced} by a root project if any of its fields, methods, or types are referenced in the root project's bytecode.

For scalability and consistency, our points-to analysis relies on static references from Java bytecode rather than dynamic traces.
Dynamic analysis can, in principle, uncover runtime-only dependencies (e.g., reflection~\cite{reflection} or configuration-driven class loading), providing complementary coverage.
However, dynamic techniques are heavy and depend on test coverage and execution scenarios, which may miss relevant paths.
Static analysis, in contrast, offers a scalable approximation of the dependency footprint, though it may introduce false positives (e.g., unused imports) and false negatives (e.g., missed reflective calls).
Given our large-scale dataset and the need for reproducibility, we focus on static analysis, following established best practices~\cite{jlbp}.

\label{sec:reachability}
\textbf{Limitations of existing techniques.}
The \texttt{dependency:\allowbreak analyze} plug-in that analyzes the undeclared used dependencies does not work on the $MCR_{root}$ dataset due to the lack of repositories and failed at 575 out of 972 modules (59\%) in $GH_{root}$, rendering it unusable. Thus, we implemented our own points-to analysis module for this study, which we claim not to be novel but to be a workable instantiation of the detection approach.
Existing static analysis techniques, such as call graph-based reachability analysis~\cite{wu2023understanding, he2024cfl, xu2024iterative, mues2024exploring, boichut2024sat, gao2024unit}, cannot measure dependency usage comprehensively. They mainly capture method invocations, but overlook field references and type dependencies.

To identify implicit dependencies, we implemented a lightweight points-to analysis on Java bytecode using SootUp~\cite{sootup}. 
Our procedure operates as follows: 
(1) we parse the compiled classes of each root project and extract all referenced \textbf{program entities}, including class types, method invocations, and field accesses;
(2) we construct a symbol table by using the Maven's \emph{effective} POM and downloading the corresponding JAR files of all direct and transitive dependencies in the resolved Maven classpath, extracting classes, methods, and fields from each JAR, and recording the declaring GAV coordinate of each entity; A dependency declared only in a parent POM is resolved to the same Level-1 position as an explicitly declared direct dependency (Section~\ref{sec:termdef})
(3) we map each referenced entity from the root project to its declaring dependency. When multiple dependencies contain classes with the same fully qualified name (a phenomenon known as dependency shadowing), we follow Maven's classpath ordering, which mirrors the classloader behavior at runtime, assigning the entity to the first dependency on the classpath that provides it;
(4) if the entity is provided only by a transitive dependency (i.e., not listed as a direct dependency in the root project's \texttt{pom.xml}), we classify the corresponding library as an implicit dependency.

\subsubsection{\textbf{Validation of the Implicit Dependency Detection Approach}}
\label{sec:validation}

Since every finding in this paper depends on our custom points-to analysis, we manually validate the tools' results by drawing a stratified random sample of 100 cases, proportional to dependency level (Figure~\ref{fig:lvl}) and $MCR_{root}$ dataset. We independently re-verify each case against artifacts by checking that: (i) the implicit dependency is absent from the root project's POM files and resolved parent chain, and (ii) the root project's decompiled bytecode references at least one class provided by the implicit dependency, as detected by scanning JVM constant-pool entries and type descriptors for every class in the root JAR. A case is manually confirmed when both conditions hold.

Two cases were excluded because their in-development artifacts were not fetchable from Maven Central. Among the remaining 98, one false positive is in fact a direct dependency declared through the parent chain that our analysis missed. Because no ground truth is available, we cannot soundly estimate recall. During manual scrutiny, we came across two sites where the used classes are not reported by our tool due to misinterpreted results from SootUp, but the dependencies were still reported based on other detected sites. We have also tried the runtime auxiliary evaluation, but it is limited by test coverage and may over-capture library-internal classes, preventing it from serving as a clean oracle. 
Manual validation cannot identify the false positives from (1) a reference that sits in a never-executed path, and (2) a class shadowed by an identically-named type in a declared dependency. 

\subsubsection{\textbf{Analysis Results for MCR}}
\label{sec:mcrresults}

To quantify the prevalence of implicit dependencies in the MCR, we analyze two key aspects: the \textbf{distribution} of occurrences of implicit dependencies across libraries and their \textbf{chronological evolution} over time.

\noindent\textbf{Distribution Analysis}:
In total, at the version level, we analyzed 19,812 versions from 1,157 unique $MCR_{root}$, among which 6,761 versions (34.12\%) contained at least one implicit dependency. At the library level, 508 out of 1,157 libraries (\textbf{43.90\%}) exhibited implicit dependencies, which indicates a significant usage of implicit dependencies. A total of 2,856 unique implicit dependencies were used by the $MCR_{root}$ libraries.
Among libraries that contain at least one implicit dependency, the median number of implicit dependencies per library version is 3 and the maximum is 87, indicating considerable variability across projects.
Moreover, the versions of implicit dependencies may change with each build over time due to the volatility of transitive dependency versions. 
Throughout our experimental period, we observed 13,842 unique implicit dependency versions (24,706 occurrences in total), with an average of 4.85 versions per dependency, highlighting significant version variability. 

\begin{figure}[t!]
    \centering
  \includegraphics[width=0.99\linewidth]{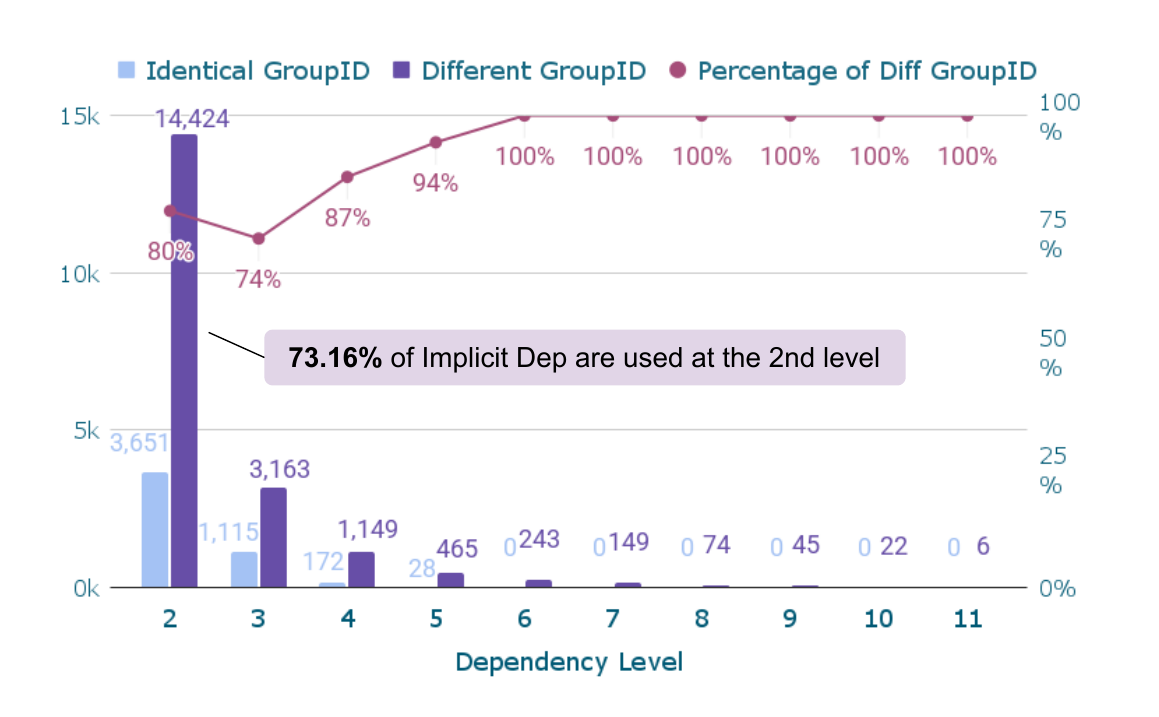}
  \vspace{-10pt}
  \caption{Implicit dependency Distribution Over Dependency Levels}
  \label{fig:lvl}
\end{figure}

The occurrences of implicit dependencies within this experiment cover 2-11 levels of transitive dependencies (Level-1 dependencies are direct dependencies). As indicated in Figure~\ref{fig:lvl}, 73.16\% of implicit dependencies are used at the second level of transitive dependencies.
This is reasonable because upper-level (lower-numbered) dependencies are more visible and often mistaken for direct dependencies, a problem exacerbated by poor API design such as unclean return types (i.e., public methods returning types defined in transitive rather than direct dependencies)~\cite{jlbp-2}.
The implicit dependencies exist at all levels, resulting in a long-tail effect. This underscores the need for comprehensive implicit dependency scanning and governance across all dependencies.

\textbf{Implicit dependencies with identical group IDs with root projects:} Libraries that share the same group ID are typically developed, versioned, and released in a coordinated manner by the same organization within a unified repository. 
Therefore, when such libraries appear as implicit dependencies, their impact is often more limited compared to libraries with different group IDs (non-native cases), since their versions are managed consistently under the same release cycle and organizational control.~\cite{dann2023upcy}
We refer to these as Native implicit dependencies, and report their distribution separately to distinguish their prevalence and risks from non-native cases.
Across all analyzed projects, 475 implicit dependency libraries (16.63\%) share identical groupIDs, indicating that only a small portion of the improper usage of transitive dependencies falls within this limited-impact category. The non-native implicit dependencies, as the majority, pose a greater risk in terms of dependency conflicts, maintainability, and unexpected behavior.

\mybox{
\textbf{Finding 1}: Among 19,812 versions, 34.12\% contained implicit dependencies, and 43.90\% of 1,157 $MCR_{root}$ exhibited implicit dependencies. A total of 2,856 dependency libraries were identified as implicit dependencies, with 16.63\% under the same groupID, meaning only a few native implicit dependencies published by the same vendor as root projects.
}

\noindent\textbf{Prevalence by popularity tier.} To test whether implicit dependencies concentrate among less popular libraries, we rank $MCR_{root}$ by Maven Central usage (the displayed ``used by'' count) and report prevalence per tier. As Table~\ref{tab:popularity} shows, prevalence is not uniform: it rises monotonically as popularity falls, from 16.0\% of the top-100 most-used libraries to 48.9\% of the least-used tier. The phenomenon is therefore concentrated among less popular libraries, yet one in six of the very most popular libraries still carries an implicit dependency, suggesting that more popular, better-maintained libraries are more likely to notice and resolve them.

\begin{table}[t]
\small

\caption{Implicit dependency prevalence by library popularity tier (Maven Central usage, reconstructed via deps.dev~\cite{osvinsight} direct dependents)}
\label{tab:popularity}
\resizebox{\columnwidth}{!}{%
\begin{tabular}{@{}lrrr@{}}
\toprule
\rowcolor[HTML]{EFEFEF}
\textbf{Popularity tier} & \textbf{\# Libs} & \textbf{\% libs w/ impl.\ dep.} & \textbf{\% vers w/ impl.\ dep.} \\ \midrule
Top 1--100    & 100 & 16.0\% & 11.0\% \\
\rowcolor[HTML]{EFEFEF} Top 101--500  & 400 & 39.5\% & 32.2\% \\
Top 501--1,156 & 656 & 48.9\% & 37.8\% \\
\bottomrule
\end{tabular}%
}

\end{table}

\subsubsection{\textbf{Analysis Results for GitHub Repos}}
\label{sec:ghresults}
In \( GH_{root} \) from GitHub, 109 out of 972 modules exhibited implicit dependencies, capturing a total of 310 implicit dependency libraries. Comparatively, the proportion of implicit dependency libraries in \( GH_{root} \) (11.21\%) is significantly lower than \( MCR_{root} \) from MCR (34.12\%). 
Upon further investigation, we found that many dependencies in \( GH_{root} \) could not be fully resolved, and some JAR files failed to download due to private or in-development versions. 
These factors reduced the coverage of our analysis and likely led to an underestimation of implicit dependency prevalence in GitHub projects. 
Nevertheless, even with this limitation, we still identified a non-trivial number of implicit dependencies, confirming that the phenomenon is not limited to published artifacts in Maven Central but also arises in actively developed projects.

\subsubsection{\textbf{Timeline Analysis}}
\label{sec:timeline}

In this experiment, we analyze the 508 root projects that exhibited implicit dependencies over their version release history to uncover the trends of implicit dependencies over time. By examining the chronological evolution of these dependencies, we aim to understand whether their occurrence is increasing, stabilizing, or declining, and how dependency management practices have evolved across different library versions.

\begin{figure}[t!]
    \centering
  \includegraphics[width=0.8\linewidth]{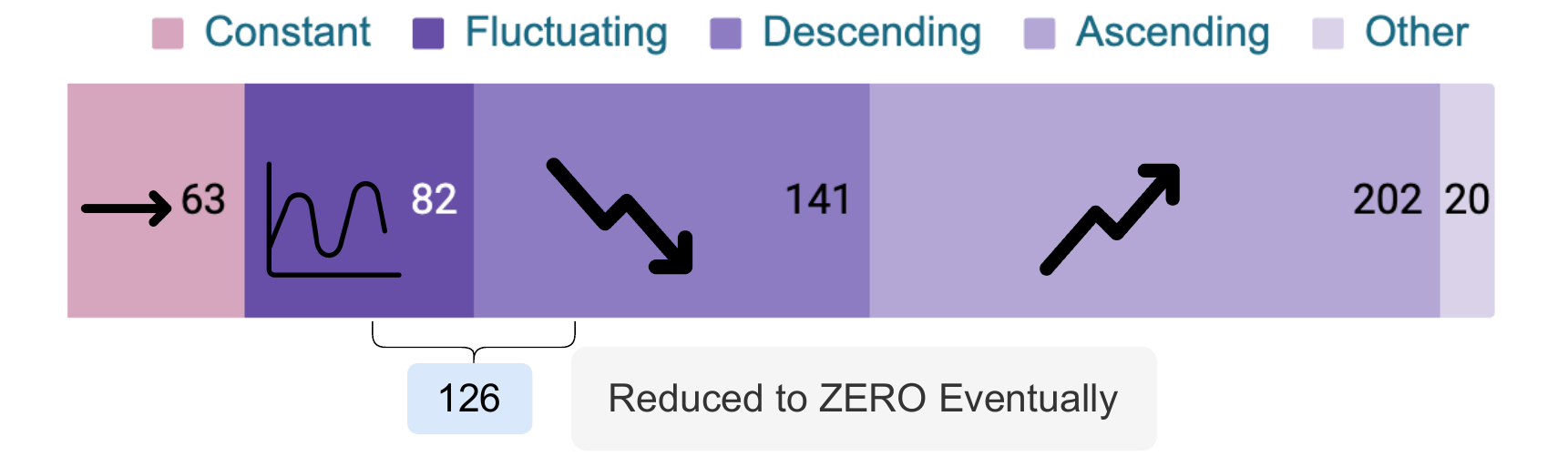}
  \vspace{-10pt}
  \caption{Trends over Time for implicit dependencies}
  \label{fig:timeline}
\end{figure}

In \Cref{fig:timeline}, the numbers of root projects regarding the trends of implicit dependencies are depicted. The trends have been categorized into five: (1) \textbf{Constant}: Implicit dependencies never change; (2) \textbf{Fluctuating}: Implicit dependency stays the same at the beginning and the end but fluctuates in between; (3) \textbf{Descending}: Implicit dependencies at the latest version are lower than the initial version; (4) \textbf{Ascending}: Implicit dependencies at the latest version are higher than the initial version; (5) \textbf{Other}: Either the versions could not be sorted by Semantic Versioning or versions of $MCR_{root}$ are too few to sort.
It is observed that 202 root libraries (39.7\%) have their implicit dependencies increased over the releases, indicating that implicit dependencies often persist and even grow across versions, suggesting that many projects continue to rely on implicit dependencies over time.

\modify{To identify how many root library developers are aware of implicit dependencies and ultimately resolve these dependencies,}
we filtered root projects to those that had implicit dependencies initially and resolved all of them in the latest releases. In Figure~\ref{fig:timeline}, only 126 (24.8\%) root projects had zero implicit dependencies in their latest versions, indicating the majority of developers failed to resolve the implicit dependencies. This experiment suggests that implicit dependencies may not be widely recognized as an issue, and there is no clear indication that its prevalence is decreasing. 

\mybox{\textbf{Finding 2}: Of the analyzed root projects, 39.7\% experienced an increase in implicit dependencies over time, while only 24.8\% of root projects reduced their implicit dependencies to zero in their latest versions. This indicates that the majority of implicit dependencies have not been eliminated, and there is no clear trend suggesting their removal in the future.
}

\subsubsection{\textbf{Characterizing Where Implicit Dependencies Occur}}
\label{sec:reason}
This analysis characterizes which kinds of libraries implicit dependencies concentrate in, as our data cannot directly evidence developer intent.
Nevertheless, we can still infer the factors that are likely to lead to implicit dependencies. A key step in this analysis is to examine the functionality of frequent implicit dependency libraries, providing insight into common characteristics of them.

\begin{figure}[t!]
    \centering
  \includegraphics[width=0.8\linewidth]{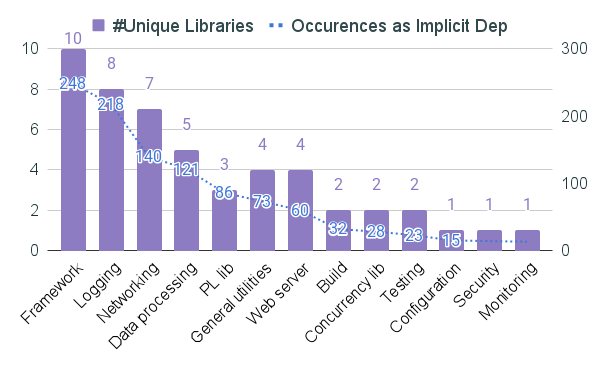}
  \caption{Categories of Top-used Implicit Dependencies}
  \label{fig:occurence}
\end{figure}

We rank the most frequently used implicit dependency libraries in descending order by combining results from both MCR and GitHub. To focus on widely recurring dependencies, we consider only those libraries with 10+ occurrences, yielding 50 unique Maven libraries. The first three authors then categorize these libraries into functional labels using a hybrid card sorting method~\cite{hybridcardsorting}. Specifically, the three authors, each with more than six years of OSS experience, independently assign tags to the libraries using self-enumerated cards, while also referring to the official Maven tags. The individual categorizations are then consolidated through discussion, during which similar categories are merged. When disagreements remain, majority voting is used to determine the final label. This process results in 13 distinct functional labels, as shown in Figure~\ref{fig:occurence}.

Among these labels, \textit{Framework} refers to the widely used Spring Framework~\cite{spring}, while \textit{PL lib} denotes programming language-native libraries other than Java, such as Scala~\cite{scala} and Kotlin~\cite{kotlin}. As shown in Figure~\ref{fig:occurence}, the most frequently occurring implicit dependency libraries originate from the Spring Framework, which consists of multiple major components such as Core, Boot, Data, and Web, each further subdivided into multiple modules. A key challenge is that these components are often used collectively, increasing the likelihood that developers may inadvertently utilize program entities directly from underlying transitive dependencies without being aware of their origin.

We compared the distribution of non-implicit transitive dependencies and found a low Pearson correlation (0.186)~\cite{pearsoncorrelation}, indicating that implicit dependencies exhibit distinct patterns from the overall trend.

\mybox{\textbf{Finding 3}: First, frameworks with numerous interdependent components, such as Spring Framework, introduce multiple associated dependency packages that are often used collectively, increasing the likelihood of implicit usage. Second, widely used libraries providing common functionalities, such as Log4j, are frequently introduced as transitive dependencies, leading to unintentional implicit usage.
}

\subsection{RQ2: Version Drift Analysis}
\label{sec:versiondrift}
We quantify how much implicit dependencies drift over time. Because they rely on soft version constraints~\cite{mavensoft,zhang2023mitigating}, their versions shift passively whenever a direct dependency is upgraded (Section~\ref{sec:versionresolution}), potentially introducing breaking changes. Even though an unpinned declared dependency drifts too, that drift is under the developer's control, following an explicit choice not to pin the version. An implicit dependency, by contrast, is absent from the manifest and shifts with no signal that the dependency even exists, let alone changed. This unaware drift, and the breaking changes it silently introduces, is what is specific to implicit dependencies. For this reason, and unlike the version-agnostic CVE rate that we compare across dependency categories in Section~\ref{sec:vulncompare}, a per-category breaking-change rate would not be directly comparable, since a declared dependency's version changes are developer-initiated whereas an implicit dependency's occur passively; we therefore characterize breaking-change risk specifically for implicit dependencies, where the drift is unaware. Note that modern package managers, such as NPM~\cite{npm} and Go Modules~\cite{gomodule}, employ lock files to ensure the stability and reproducibility of dependency versions, but this does not eliminate the risk of implicit version drift, as lock files are typically generated and updated automatically when the direct dependencies are modified and re-locked.

Since implicit dependency libraries drift differently under different root projects, our analysis is conducted in pairs consisting of a root project and an implicit dependency, resulting in a total of 2,856 pairs for evaluation. 
To quantify the extent of version drift, we first measure the version spans of implicit dependency pairs. Additionally, we calculate the number of major version crossings of implicit dependencies, because, according to Semantic Versioning~\cite{semver}, major version changes permit breaking changes, thus introducing instability. 

\subsubsection{\textbf{Version Span of Implicit Dependencies}}

The version span of an implicit dependency refers to the number of versions between the initial version and the changed version (version span is 1 for two consecutive versions), reflecting the extent of variation in the versions of implicit dependencies. A larger version span indicates greater instability and increased risk for the project. Specifically, the old version denotes the first version of \(MCR_{root}\) in which the implicit dependency is introduced, while the new version denotes either the version immediately before its removal from \(MCR_{root}\), or the latest available version of \(MCR_{root}\) if it has not been removed.

To measure the version span, we first retrieved the complete version list of the implicit dependencies from MCR. The versions were then sorted in ascending order using Maven's official semantic versioning tool~\cite{mvngithub}. Finally, we calculated the number of versions in between to quantify the evolution of the implicit dependencies.

The median version span across all implicit dependency pairs is 7 (mean: 36.16), indicating that version spans in implicit dependencies are both common and non-trivial.
Some artifacts, such as \texttt{aws-core}~\cite{awscore}, have released over a thousand versions, resulting in outliers that inflate the overall distribution. To address this, we applied Tukey's method~\cite{tukey} to exclude outliers by removing 315 implicit dependencies with 100+ version spans. For the remaining 2,856 implicit dependency pairs, 44.15\% of implicit dependencies have 10+ version spans. Such spans are still likely to introduce breaking changes and unexpected behaviors~\cite{jayasuriya2023understanding}, posing potential risks to the stability of the root projects
\cite{ochoa2021breaking,jayasuriya2023understanding,raemaekers2017semantic}.

Regarding major version crossing, it turned out 18.45\% of implicit dependencies exhibit major version crossings in their version spans, indicating a high likelihood that projects depending on these implicit dependencies have faced unexpected breaking changes. Even for non-major version crossing, there remains a probability of introducing breaking changes, as supported by prior studies~\cite{ochoa2021breaking, jayasuriya2023understanding, raemaekers2017semantic}.

It is important to note that these figures may significantly underestimate the actual impact of implicit version drift in real-world development. This is because \( MCR_{root} \) are typically in a stable state when published on MCR, either as compilable and tested source code or as off-the-shelf JAR files. Any unexpected issues encountered during development, including those caused by implicit dependencies, are likely to have been addressed prior to the publishing. 

\mybox{\textbf{Finding 4}: Over 44.15\% of implicit dependencies span more than 10 versions throughout their lifetime, and 18.45\% experience major version changes. It indicates a high likelihood of breaking changes and unexpected behaviors caused by untracked version variations in directly used implicit dependencies.}

\subsubsection{\textbf{Code-Centric Breaking Change Analysis}}
\label{sec:breakingcode}
To quantitatively analyze the evolution of breaking changes in implicit dependencies caused by version drift, we aim to measure the number/proportion of broken program entities (such as classes, methods, and fields) introduced by these version changes. This metric has been commonly used to measure the level of compatibility by research works~\cite{ponta2018beyond,pashchenko2020vuln4real,steady,ponta2020detection}. This analysis is conducted through a code-centric examination of the JAR files associated with both the root projects (\( GH_{root} \) and \( MCR_{root} \)) and implicit dependencies. 
Based on the points-to analysis introduced in Section~\ref{sec:method}, we identify the breaking APIs and constructs that are utilized by the root projects. Furthermore, to assess the extent of effort invested in mitigating breaking changes, we examine which program entities have ceased to be used as a response to these disruptions.

Specifically, we measure breaking program entities by utilizing \texttt{revapi}~\cite{revapi}, a widely adopted tool for detecting differences between JAR files, which has been extensively used in prior research~\cite{ochoa2021breaking, jayasuriya2023understanding, zhang2023mitigating, zhang2023compatible}. Combined with our points-to analysis, this approach allows us to precisely identify the program entities affected by breaking changes.
However, reproducing these breaking changes in practice is challenging, as developers must have already addressed and resolved them before publishing to MCR and GitHub. Thus, it is unrealistic to reproduce the breaking change with published artifacts. This inherent limitation suggests that the observed breaking changes represent only those that persisted through development and were not mitigated prior to releases.

\begin{figure}[t]
  \centering
  \includegraphics[width=0.7\linewidth]{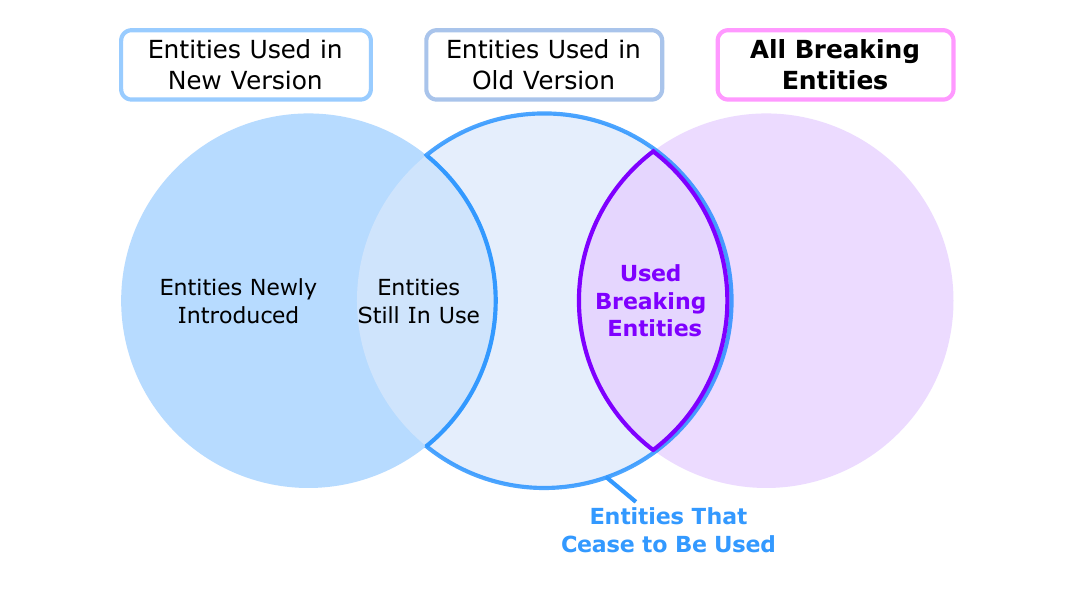}
  \caption{Venn Diagram of \textbf{Used} Entities in implicit dependencies}
  \label{fig:venn}
\end{figure}

\begin{table}[]
\setlength{\tabcolsep}{2.5pt}
\footnotesize
\caption{Distributions of 2,856 implicit dependencies with breaking changes and Associated Breaking Entities}
\resizebox{\columnwidth}{!}{%
\begin{tabular}{@{}l|rrr@{}}
\toprule
\rowcolor[HTML]{EFEFEF}
\textbf{}                     & \multicolumn{1}{l}{\cellcolor[HTML]{EFEFEF}\textbf{Potentially Breaking}} & \multicolumn{1}{l}{\cellcolor[HTML]{EFEFEF}{\color[HTML]{333333} \textbf{Used Breaking}}} & \multicolumn{1}{l}{\cellcolor[HTML]{EFEFEF}\textbf{Cease to Use}} \\ \midrule
 \# implicit dependencies& 1,373 (48.07\%)                                                    & 649 (22.72\%)                                                                             & 1,134 (39.71\%)                                                    \\
\rowcolor[HTML]{EFEFEF} \# Breaking-change entities                      & 265,074                                                           & 27,573                                                                                    & 62,179                                                            \\
 \bottomrule
\end{tabular}%
}
\label{tab:breaking}
\end{table}

\textbf{Breakdown of Entities in implicit dependencies}: Figure~\ref{fig:venn} illustrates how used entities in an implicit dependency evolve when the dependency drifts from version $v_1$ to $v_2$. Some entities continue to be used, some new ones are adopted, and some are abandoned. Among the \textit{Entities that cease to be used}, some may be breaking entities that previously caused compatibility issues. For published artifacts, such breaking entities are often no longer used because root projects on MCR and GitHub are typically released in a compilable and functional state, although unresolved breaking changes may still occasionally remain. We therefore use this diagram to show how many breaking entities were used in the old version and how many were later resolved.

In Table~\ref{tab:breaking}, the counts and proportions of all 2,856 implicit dependencies are presented. Notably, approximately 48\% of implicit dependencies have undergone version drift involving breaking changes. However, not all of these have resulted in the usage of breaking entities within root projects. Our points-to analysis reveals that around 22\% of implicit dependency libraries have introduced breaking entities actually used in the old version to root projects through version drift.
This suggests that many developers likely encountered unexpected breaking changes as their projects evolved and direct dependencies were upgraded, but resolved them before release. This interpretation is supported by the fact that the number of entities that ceased to be used exceeds the number of used breaking entities, since it also includes abandoned non-breaking entities. Interestingly, only 39\% of implicit dependency libraries were entirely removed from usage, indicating that many potentially breaking entities remain in use.

\mybox{
\textbf{Finding 5}: Approximately 48\% of implicit dependencies have introduced breaking changes due to version drift driven by the evolution of root projects, and 22\% of them have their breaking entities actively used by the root projects. This indicates that a significant number of developers may have encountered breaking issues during development, causing a substantial impact. 
}

\subsection{RQ3: Security Issues Evolution in Implicit Dependencies}
\label{sec:vuln}

This section investigates security risks in implicit dependencies, which are critical because they may be overlooked by widely used SCA tools such as Dependabot~\cite{dependabot} and OWASP Dependency-Check~\cite{owaspcomponent}. For example, Dependabot primarily reports and recommends upgrades for direct dependencies~\cite{dpb}, although it can scan transitives when a lock file is available. However, prior studies~\cite{wu2023understanding,zhang2023mitigating,hu2019open} show that vulnerabilities also reside in transitive dependencies, and vulnerabilities in implicit dependencies are especially easy to overlook compared to those in other transitive dependencies, because, even when reported as vulnerable transitives, their direct usage by root projects can make their impact easy to underestimate.
To assess these risks, we examine Common Vulnerabilities and Exposures (CVE) associated with implicit dependencies to raise attention on their potential security implications and promote the actions to resolve them timely.

\subsubsection{\textbf{Vulnerability and Data Collection}}
\label{sec:vulndata}
The vulnerability mapping between dependency libraries and CVE entries from the National Vulnerability Database (NVD)~\cite{nvd} was previously collected and stored in a database for querying. Specifically, the mapping of each library was determined by aligning the CPE (Common Platform Enumeration)~\cite{cpe} fields of CVE entries with Maven library names, supplemented by manual validation to ensure accuracy.
The version of each implicit dependency used for this range check is the version resolved by standard Maven semantics as described in Section~\ref{sec:mcr}: the root release's own \texttt{pom.xml} constraints where tightly pinned, and the May-2025 Maven Central state for soft-constraint mediation otherwise. For tightly pinned constraints this matches the version that was actually resolved at the root project's release time; for soft constraints, the version we check against a CVE's range may reflect the May-2025 state rather than the exact version resolved at the root project's original release date, which we note as a threat to validity (Section~\ref{sec:internalvalidity}).
As a result, we collected CVE mappings from NVD for 909 unique implicit dependency libraries, covering a total of 8,271 associated versions. Using the version ranges specified in CVE entries, we mapped the corresponding versions of implicit dependencies to assess their security impact. In total, we have collected 692 CVEs.

\subsubsection{\textbf{Analysis of Vulnerabilities}}

The statistical analysis reveals that 865 implicit dependency libraries (30.28\%) and 2,179 library-version pairs are affected by CVEs. On average, each affected implicit dependency library is associated with 5.78 CVEs, highlighting an upper bound of security risk for implicit dependencies.
To validate the real security risks, we extracted CVE-affected methods and cross-checked against the used methods in implicit dependencies via our points-to analysis. Among the 692 identified CVEs, $445$ were associated with vulnerable methods. Cross-checking these with the used methods in implicit dependencies, we manually confirmed that \textbf{36} CVEs were actually used by root projects, highlighting the tangible security threats that implicit dependencies introduce. 

\begin{figure}[t!]
  \centering
  \includegraphics[width=0.70\linewidth]{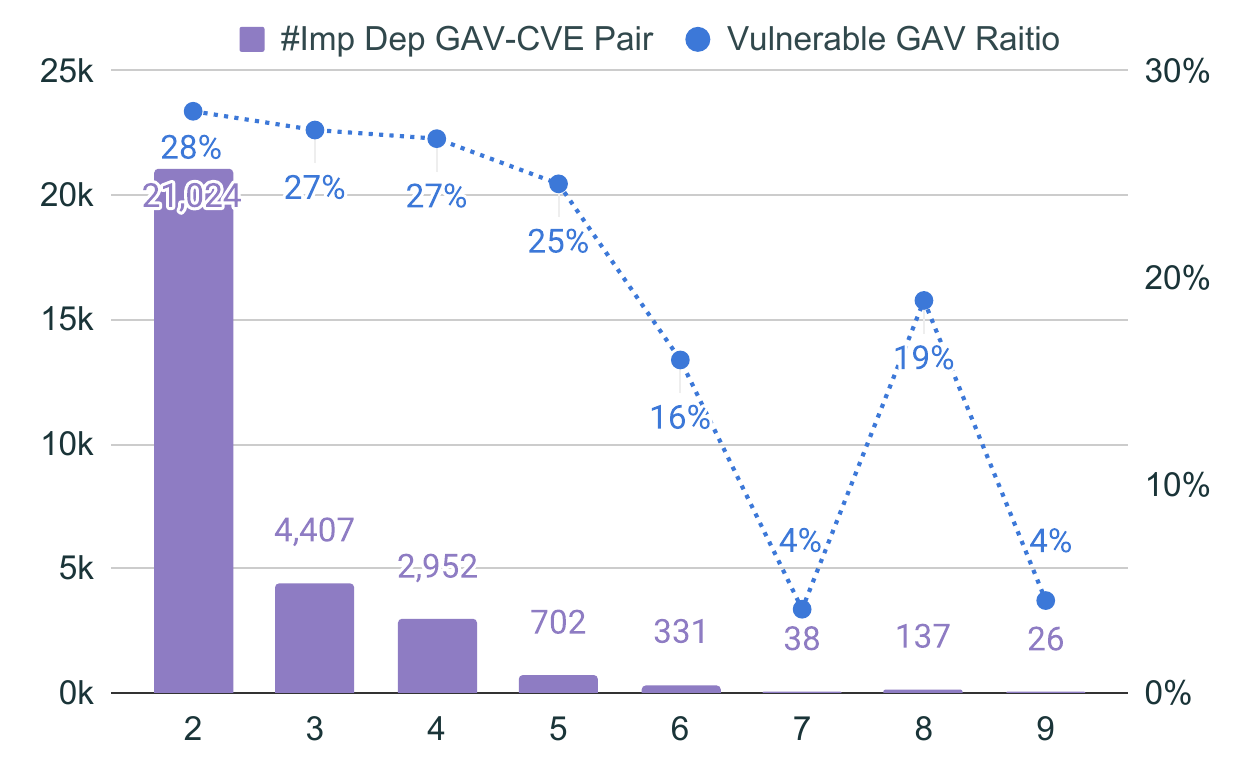}
  \caption{Vulnerabilities Distributions per Level}
  \label{fig:vullvl}
\end{figure}

We then measured the security risks by \textit{CVE counts} based on one combination of a root project, a version (GAV) of implicit dependencies, and a CVE occurrence, representing an instance where a vulnerability is found.
The sum of these CVE counts per dependency level is plotted in Figure~\ref{fig:vullvl}, where a generally decreasing trend is observed. The CVE count at level 2 is much higher than at other levels, primarily because more implicit dependencies appear there. However, the proportion of vulnerable GAVs at level 2 is not significantly higher than at other levels; despite slight fluctuations, it remains relatively stable at around 20--30\% across levels. This indicates a non-trivial security risk that does not diminish with dependency depth. Because implicit dependencies are directly referenced by root projects, vulnerable code within them is more likely to be executed than in unused transitives. Although our static analysis cannot establish exploitability with certainty, it still suggests elevated risk; a more precise assessment would require call-graph or dynamic analyses (e.g., \cite{hejderup2022prazi, pashchenko2020vuln4real}).

\subsubsection{Comparison of Vulnerability Prevalence among Dependency Categories}
\label{sec:vulncompare}
Unlike the unaware version drift of Section~\ref{sec:versiondrift}, vulnerability incidence is not specific to implicit dependencies, but a general property of any dependency. To situate this figure against other dependency categories, we re-measured CVE rates on the $MCR_{root}$ across three categories: declared direct dependencies, implicit dependencies, and other (non-implicit) transitive dependencies, using a single consistent vulnerability source (OSV~\cite{osv}) so the rates are directly comparable. As Table~\ref{tab:vulncompare} shows, implicit dependencies (23.4\%) are affected at a rate comparable to, and marginally below, declared direct dependencies (27.6\%), with both exceeding unused transitive dependencies (14.2\%). Implicit dependencies are therefore not more likely to be vulnerable than the dependencies developers already audit; a known-vulnerability rate is a general property of reused code. What distinguishes the implicit subset is not the rate but the auditing gap: because SCA tools prioritize declared dependencies, their higher CVE rate is a structural blind spot for the implicit subset.

\begin{table}[t]
\small

\caption{CVE incidence by dependency category on the same $MCR_{root}$ sample (OSV, 60 roots)}
\label{tab:vulncompare}
\begin{tabular}{@{}lrr@{}}
\toprule
\rowcolor[HTML]{EFEFEF}
\textbf{Dependency category} & \textbf{\# instances} & \textbf{\% with $\geq$1 CVE} \\ \midrule
Declared direct            & 283 & 27.6\% \\
\rowcolor[HTML]{EFEFEF} Implicit (used, undeclared) & 47  & 23.4\% \\
Other transitive (unused)  & 316 & 14.2\% \\
\bottomrule
\end{tabular}

\end{table}

\mybox{\textbf{Finding 6}:
36 CVEs have vulnerable methods directly used by root projects. Under the version-range convention SCA tools apply to declared dependencies, 30.28\% of implicit dependencies are affected by known vulnerabilities.
}

\subsection{RQ4: Observed Resolution Outcomes for Implicit Dependencies}
\label{sec:countermeasure}
In this research question, we characterize how implicit dependencies disappear from root projects over successive releases.
First, we estimate the developers' awareness of implicit dependencies by evaluating their exposure to breaking changes caused by version drift. These breaking changes are evident issues that must be resolved before publishing.
Subsequently, we analyze how implicit dependencies disappear over releases by tracking the evolution of \texttt{pom.xml} files and examining code-level modifications in root projects in which implicit dependencies are no longer detected.

\subsubsection{\textbf{Statistics of Resolution of Implicit Dependency Libraries}}
To statistically analyze developers' reactions and countermeasures toward implicit dependencies, we first examine how many implicit dependencies have been removed throughout the development lifecycle of root projects.
Note that we have used keyword matching to locate related commits in $GH_{root}$, using terms such as ``implicit dependency'', ``undeclared dependency'', ``ghost dependency'', ``transitive dependency fix'', and ``add missing dependency'' along with their common synonyms, but have found nothing relevant (commits on our website~\cite{dataset}).
Although there is no explicit indication that developers intentionally fix implicit dependencies, analyzing their disappearance provides insights into their overall reduction.
Focusing on root projects with multiple version releases: if an implicit dependency appears in a project and ceases to be an implicit dependency later, regardless of how they are resolved, this occurrence is recorded. To avoid misclassifications, the record does not count if the same implicit dependency re-emerges in subsequent versions of the root projects.

Our findings reveal that out of 2,856 analyzed dependencies, 1,039 instances of root-project-dependency pairs have resolved implicit dependencies, representing 36.38\%. It also suggests that many implicit dependencies remain unresolved and continue to persist.

To analyze how many developers fix implicit dependencies intentionally due to breaking changes, we refer to the previously derived statistics in Section~\ref{sec:breakingcode}, where our analysis in Table~\ref{tab:breaking} indicates that 22.72\% of implicit dependency libraries have experienced breaking changes.
Among the 1,039 resolved implicit dependency libraries, 97.37\% belong to the set of 1,373 implicit dependencies with breaking changes, confirming that approximately 22\% of implicit dependencies were likely removed because root project developers encountered breaking changes due to version drift. The slight discrepancy in overlap (97.37\% instead of 100\%) may be attributed to occasional inaccuracies in our breaking change prediction methods.

Showing intent requires commit-message evidence that developers recognize and deliberately fix implicit dependencies. To probe this beyond our dataset, we ran GitHub commit-message searches for implicit-dependency terms (e.g., ``used undeclared dependency''), yielding 235 candidate commits. In a manually inspected sample of 100, 65 were genuine dependency-declaration commits across 48 projects and five ecosystems (Maven, Gradle, Composer, npm, PyPI); most quote a dependency-analysis tool. Thus developers do occasionally resolve implicit dependencies deliberately, but almost always only after a tool surfaces them. This also underscores our solution's potential to reveal many more implicit dependencies and contribute to better dependency management practices.

\mybox{\textbf{Finding 7}: Among 2,856 analyzed implicit dependency libraries, 1,039 were resolved, and about 22\% were associated with breaking changes introduced by version drift.
}

\subsubsection{\textbf{Observed Resolution Patterns}}
We further investigated the resolution patterns observed when an implicit dependency disappears, by analyzing changes in \texttt{pom.xml} files and related code modifications in newer versions of root projects. Note that developers may not have taken these measures intentionally, as our analysis is based on observed outcomes rather than developer decision-making. Therefore, this analysis aims to unveil potential solutions that could be leveraged to fix implicit dependencies, rather than attributing deliberate intent to developers' actions.
We explored the potential strategies and categorized the patterns and found that 1,039 cases could fall into these categories, including:

\begin{itemize}[leftmargin=1.5em]
    \item \textbf{Promotion to Direct Dependencies} (160 cases, 15.40\%): The implicit dependency is declared as a direct dependency in \texttt{pom.xml}.
    \item \textbf{Evolutionary Removal} (158 cases, 15.21\%): The implicit dependency has evolved out of the dependency tree, meaning that while the direct dependencies remain (albeit possibly with version updates), implicit dependency is no longer a dependency.
    \item \textbf{Removal of Direct Dependents} (207 cases, 19.92\%): The direct dependencies that previously introduced implicit dependencies are removed, causing implicit dependencies to disappear.
    \item \textbf{No Longer Used but Still Present} (514 cases, 49.47\%): The implicit dependency remains in the dependency graph but is no longer referenced in the root project’s code, effectively becoming a normal transitive dependency without direct usage.
\end{itemize}

The most common resolution (49.47\%) is that implicit dependencies simply cease to be referenced in the root project's code, suggesting that many developers may not be aware of implicit dependencies and instead modify code for other reasons that incidentally resolve the issue.
In contrast, promotion to direct dependencies (15.40\%) is the only strategy that requires no code-level modifications---only updating \texttt{pom.xml}---and inherently reflects developer awareness.
The remaining strategies (evolutionary removal and removal of direct dependents) likely involve developer intervention but may not have been undertaken explicitly to address implicit dependencies.

\mybox{
\textbf{Finding 8}: Promoting implicit dependencies to direct dependencies is the least effortful but requires developer awareness, while other strategies involve more costly code changes and may not reflect such awareness. The most applied way is that developers simply stop using implicit dependency libraries but leave them in the project.
}

\section{Discussions}

\subsection{Threats to Validity and Limitations}

\subsubsection{Internal Validity}
\label{sec:internalvalidity}

\noindent\textbf{Static analysis accuracy.}
Our points-to analysis may still introduce false positives (e.g., bytecode references never exercised at runtime) and false negatives (e.g., reflection or configuration-driven loading). We nevertheless chose static analysis because it scales to 19,812 versions, is reproducible, and does not depend on test coverage, which is unavailable for many published JARs~\cite{smaragdakis2015pointer, zhu2004symbolic, milanova2004precise, steady, zhang2023compatible, zhang2023mitigating, snyk}. We archived intermediate outputs to improve transparency.

\noindent\textbf{Dependency shadowing.}
When multiple dependencies provide the same fully qualified class name, we resolve entities by Maven classpath order, matching standard classloader behavior. Custom classloaders may therefore cause occasional misattribution, although duplicate classes are uncommon in well-maintained Maven projects and are unlikely to affect our prevalence-level conclusions.

\noindent\textbf{Bounded version sampling.}
For libraries with more than 20 versions we evenly sample 20 releases, so an implicit dependency that appears and disappears entirely between two samples escapes our timeline analysis. We assess this directly by re-detecting on the full release history of 31 high-version libraries (up to 106 releases, median 65; 1{,}977 versions) and recomputing our two headline conclusions both ways. Version-level prevalence is 48.6\% on the full history versus 49.6\% on the sample, and only 6 implicit dependencies across all 1{,}977 versions fall strictly between two samples. Moreover, 30.8\% of libraries have $\leq$20 releases and are analyzed exhaustively, so any missed dependency would be under- rather than over-counted; exhaustive analysis remains infeasible only for the longest histories.

\noindent\textbf{Gradle v.s. Maven.}
Our MCR analysis includes both Maven- and Gradle-published artifacts. To quantify any effect, we label a library Gradle-published when its latest release carries Gradle Module Metadata (\texttt{.module}) or the ``published-with-gradle-metadata'' marker, and Maven-published otherwise. Gradle-published artifacts account for 230 of 1{,}156 libraries (19.9\%) and 4{,}394 of 19{,}812 versions. Implicit dependencies are modestly more prevalent among them (38.10\%) than Maven-published versions (32.99\%), consistent with Gradle's generated POMs omitting some scopes. Restricting to Maven-published artifacts shifts the pooled version-level rate only from 34.12\% to 32.99\%, confirming the phenomenon holds in both groups.

\subsubsection{External Validity}

\noindent\textbf{Sampling bias.}
Our dataset prioritizes widely used Maven Central libraries and star-ranked GitHub repositories, which biases the sample toward popular and relatively well-maintained projects. We chose this deliberately because these projects have the largest downstream impact, but less visible or proprietary systems may exhibit different implicit dependency rates.

\noindent\textbf{Build system scope.}
We focus on Maven-managed Java projects because our analysis relies on \texttt{pom.xml} files and \texttt{mvn dependency:\allowbreak tree}. Maven remains the dominant publishing format for Maven Central, but Gradle-native, Ant, Bazel, and other non-Maven projects fall outside our direct scope.

\noindent\textbf{Ecosystem generalizability.}
Our study is restricted to Java/Maven and may not directly transfer to ecosystems with different lock-file practices, module systems, or typing disciplines. Still, the core phenomenon of directly using undeclared transitives is not unique to Maven as revealed by our GitHub search~Section~\ref{sec:countermeasure}.

\subsection{Implications}

\subsubsection{Raising Awareness and Automated Detection of Implicit Dependencies}
Awareness of implicit dependencies should be raised among developers and package managers to enable proactive detection during builds, with static analysis issuing warnings since early detection beats unintentional oversight. IDEs should play a more active role: neither IntelliJ IDEA~\cite{intellij} nor Eclipse with the m2e plugin~\cite{eclipsem2e} currently warns when an imported symbol resolves through a transitive rather than a declared dependency---both flag only unresolved imports, not implicit ones. Rather than only rendering dependency graphs for manual inspection, IDEs should raise automated alerts for implicit dependencies, especially for commonly used types such as logging libraries (Section~\ref{sec:reason}).

\subsubsection{Effective Resolution Strategies for Implicit Dependencies}
Section~\ref{sec:countermeasure} shows multiple resolution approaches, depending on whether developers truly need the functionality. The most straightforward is to promote implicit dependencies to direct ones with explicit version control, which package managers and IDEs could offer as a default recommendation on detection; more sophisticated strategies could add automated, context-aware code adjustments.

\subsubsection{Abandonment of Traditional Direct Dependency Definition for SCA Tools to Achieve Comprehensive Scanning}

Widely used tools (e.g., Snyk~\cite{snyk}, OWASP Dependency-Check~\cite{owasp}, Sonatype~\cite{sonatype}, JFrog Xray~\cite{jfrogxray}) already scan transitive dependencies for known vulnerabilities, but treat all transitives uniformly, ignoring whether they are actually referenced in the root project's code---exactly the directly-used-yet-undeclared subset our study isolates as a distinct, easily overlooked risk surface. Our results thus complement SCA/SBOM capabilities by motivating usage-aware scanning that scrutinizes implicit dependencies as closely as declared ones.

\section{Related Work}
\label{sec:relatedwork}

\subsection{Maven Dependency Management}

Used-but-undeclared dependencies have surfaced as one facet of broader dependency-management studies; we therefore review work on Maven dependency-management challenges---security vulnerabilities, bloated dependencies, and conflicts.

Cataldo et al.~\cite{cataldo2009software} first distinguished implicit from explicit dependencies but did not quantify their prevalence or risks.
LaLou et al.~\cite{lalou2013apache} described Maven's dependency design; Soto-Valero et al.~\cite{soto2021comprehensive} found 57\% of transitives bloated and proposed safe removal, their used-transitive label matching our phenomenon but as one of six bloat labels, without tracking breaking changes or CVEs.
Outside Maven it is only secondary: Javan Jafari et al.~\cite{javanjafari2022dependency} list a missing-dependency smell among several NPM smells, and Latendresse et al.~\cite{latendresse2022not} note but exclude transitively-resolved missing peer dependencies---neither isolates it nor targets Maven/Java.
Schott et al.~\cite{schott2026uncovering} target shaded and relocated hidden dependencies via bytecode fingerprinting, orthogonal to our focus.
Düsing and Hermann~\cite{dusing2022transitive} analyzed the security impact of direct and transitive vulnerabilities, and Jezek and Dietrich~\cite{jezek2014safeguard} built a static-analysis tool to safeguard recursive resolution and flag duplicate libraries.
Longitudinally, Zerouali et al.'s npm technical-lag analysis~\cite{zerouali2018empirical} finds versions drift from their latest release over a project's lifetime, consistent with our version-drift results (Section~\ref{sec:versiondrift}).

Several tools focus on remediation: Upcy~\cite{dann2023upcy} upgrades with minimal breaking changes, Jaime et al.~\cite{jaime2024balancing} balance update cost and quality, and Coral~\cite{zhang2023compatible} integrates downstream compatibility; Przymus et al.~\cite{przymus2025out} studied the lifecycle of transitive vulnerabilities in Maven, and Zhao et al.~\cite{zhao2023fse} evaluated SCA tools.
These works highlight the need for remediation and pruning, whereas our study focuses on the subset of transitives that are used but undeclared, showing how they persist and drift in practice.
For dependency conflicts, Wang et al. proposed Decca~\cite{wang2021will} and Riddle~\cite{wang2019could}, VeriBuild~\cite{fan2020escaping} combined static and dynamic analysis to detect build errors, and Jayasuriya et al.~\cite{jayasuriya2023understanding} examined client-impacting breaking changes.

Despite these efforts, none isolates implicit dependencies as the focal phenomenon across their lifecycle. Our work fills this gap, quantifying their prevalence, version-drift risk, and evolution.

\subsection{Human-Involved Qualitative Studies}
Qualitative studies complement ours by capturing developer perspectives. Pashchenko et al.~\cite{pashchenko2020qualitative} interviewed 25 developers and found reasoning about dependency security often challenging; Miller et al.~\cite{miller2023we} showed that most developers lack the knowledge to manage unmaintained libraries; and He et al.~\cite{he2023automating} analyzed developer experiences with Dependabot. These studies offer critical insights but are limited by small samples; our quantitative analysis complements them with ecosystem-scale evidence.

\section{Conclusion}

This study systematically investigated implicit dependencies, revealing their widespread presence, long-term persistence, and associated risks. We found that implicit dependencies often introduce breaking changes through version drift, with 22.72\% of dependencies susceptible; security vulnerabilities affect 30.28\%, yet many remain unnoticed by developers and existing SCA tools; and while 36.38\% were eventually resolved, most resolutions were unintentional, highlighting the need for proactive management. These findings call for greater awareness, better dependency management, and tooling to mitigate these risks.

\section*{Acknowledgments}
This research is supported by the National Research Foundation Singapore, Prime Minister's Office, Singapore, and the Cyber Security Agency under the National Cybersecurity R\&D Programme (NCRP25-P04-TAICeN) and its Campus for Research Excellence and Technological Enterprise (CREATE) programme.

\section*{Data Availability Statement}
Our dataset and scripts are accessible at \url{https://doi.org/10.5281/zenodo.21280403}.

\bibliographystyle{ACM-Reference-Format}
\bibliography{acmart_used}

\end{document}